# Classification of optics-free images with deep neural networks

**SOREN NELSON,[1] AND RAJESH MENON[1,*]**

[1]*Department of Electrical & Computer Engineering, University of Utah, Salt Lake City, UT 84112, USA*
**rmenon@eng.utah.edu*

**Abstract:** The thinnest possible camera is achieved by removing all optics, leaving only the image sensor. We train deep neural networks to perform multi-class detection and binary classification (with accuracy of 92%) on optics-free images without the need for anthropocentric image reconstructions. Inferencing from optics-free images has the potential for enhanced privacy and power efficiency.

There is strong interest in reducing the thickness of cameras, driven primarily by consumer electronics, [1] but also by biomedical and related applications. The vast majority of the resulting research attempts to reduce the number of, and thicknesses of the constituent optical elements. Examples of these include metalenses [2] and multi-level diffractive lenses[3]. It is also recognized that the thickness of a camera is dominated by the space between the sensor and the optics, and sophisticated methods are being proposed to reduce this, even for modest improvements [4]. However, it is clear that the ultimate reduction in camera thickness is achieved by completely removing the optics altogether. [5-7] Such optics-free cameras require computational post-processing to recover the images for human consumption, [5,6] or to recover information for machine inferencing. [6,7] In this paper, we show advances in the latter by performing classification on the raw optics-free images with trained deep neural networks. Although our classification accuracy is slightly worse than that achieved via optics-based cameras, this can be improved when the imaging context is pre-defined, a regime we previously referred to as *application-specific imaging*. An important advantage of these optics-free cameras is their inherent barrier to human perception, which has the potential to enhance privacy.

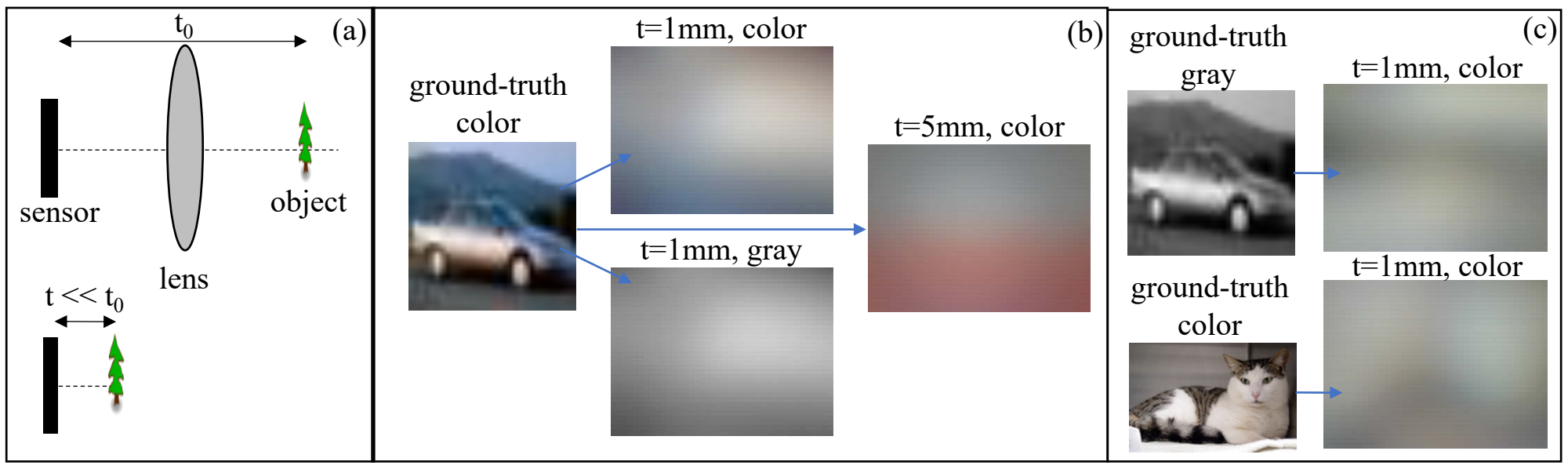

Fig. 1. (a) Optics-free cameras enable drastic reduction in thickness by removing the lens as well as decreasing the space between the sensor and the object. (b) Example of a ground-truth color image (Cifar), and raw sensor images at t=1mm (top: color and bottom: grayscale), and t=5mm. (c) Example (top) grayscale and (bottom) color ground-truth images, and corresponding color sensor images at t=1mm. The bottom image is an example from the set used for binary (cats vs dogs) classification experiments.

In conventional lens-based cameras, the smallest distance between an object and its corresponding real image ($t_0$) is limited to four times the effective focal length of the system of lenses (Fig. 1a). By forgoing image formation altogether, this distance (t) can be far smaller. In our experiments, the object is a photograph displayed on a liquid-crystal display (LCD, Acer

G276HL 1920 X 1080). The sensor (Mini-2MP-Plus, Arducam, pixel width 2.2µm) is placed at a distance t =1mm or 5mm away from the LCD. We explored both color and grayscale images in the context of two separate binary classification problems.

The first is binary classification on a *class-imbalanced dataset*. Class-imbalance problems are common in many real-life applications where the distribution of examples across classes is skewed such as in medical diagnoses, biometric identification, spam detection, video surveillance, oil spill detection, etc. Here we look at the binary case where 90% of the examples belong to a single majority class and the other 10% to a minority class. The majority class is formed by the merging of all, but one of the classes. We refer to this as a *detection problem* because the network is detecting whether or not a minority object exists in an image (for example, the truck in Fig. 1b). The second problem explores the potential of making distinctions between two similar looking images (for example, cats vs dogs, see Fig. 1c) via binary classification of fine-grained, balanced classes. In both cases, the proposed system achieves results approaching those achieved with the original images, which is remarkable considering the loss of information attributed to the optics-free "imaging" process.

The experiments were performed in a dark room except for the LCD, which was fully covered aside from the image to minimize excess light. The displayed images were all resized to 20 X 20 pixels (physical size of 6mm X 6mm) to approximately match the size of the sensor. The output of the sensor (320x240 pixels in 3 color channels) is used as input to the DNN, which is initially trained to classify the objects in the images. For the detection problem, we used images from the CIFAR-10 dataset,[8] which consists of natural color images from ten equal sized classes: 4 vehicles and 6 animals. Each image is resized from 32x32 to 20x20 (pixels). At a distance, t=1mm, the field of view is $2 \times \tan^{-1}(6/2) = 143°$. To estimate the amount of information lost in the optics-free imaging process, we computed the average absolute difference between the Shannon entropy of the ground-truth and the sensor images. The average information loss values were: 12% (t=1mm, color), 19% (t=1mm, gray display), 14% (t=1mm gray sensor), 19% (t=5mm, color), compared to only 3.7% for the lensed images. There were a total of 60,000 images in the dataset, of which 45,000 were used for training, 5,000 for validation, and the remaining 10,000 for testing. To create a detection problem, we merged 9 classes to form the negative majority class, and used the last class as the positive minority class. The ground truth label of the positive class was 1, while the negative class was 0. Thus, ten binary sets were created, one for each of the 10 classes.

A base model was pretrained to classify the images into the original 10 CIFAR-10 classes. The base model weights were then transferred to 10 binary classifiers, each of which were fine-tuned for the detection task on the corresponding dataset (see above). This pretraining allows the network to learn more general features about the dataset than the binary classifier would, when trained without the pretraining, [9] which reduces the problem of overfitting that is common when dealing with imbalanced classes. In addition, the minority class is oversampled in the training set, which has been shown to be another effective method to handle class imbalance [10]. The oversampling was done only for the binary classifier, and not the base model. An oversample rate of 5X proved empirically to give the best results.

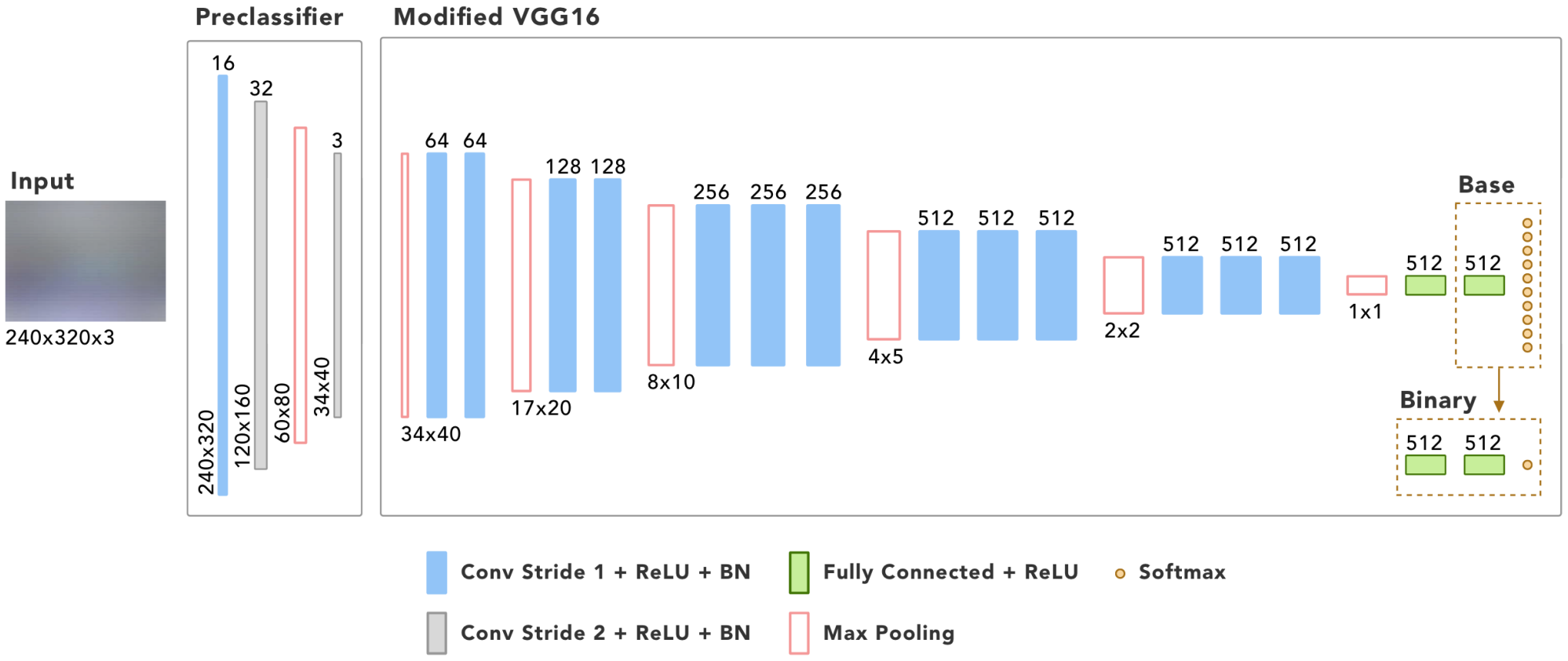

Fig. 2: Architecture of the preclassifier and the modified VGG16 network. An example input image is shown on the left.

The base model uses the VGG16 [11] architecture with slight modifications. Prior to the VGG16, the images were reduced in size by a preclassifier consisting of a max pooling and 3 convolutional layers. The subsequent VGG16 network is comprised of batch normalization and smaller fully connected layers. For the binary classifier, all network weights were transferred from the base model, except the last 2 fully connected layers, which were replaced with 3 new fully connected layers. The last layer contains a single neuron with a softmax activation. (Fig. 2). We found that the base network with the modified VGG16 architecture, and our preclassifier was able to achieve similar or better results than newer state of the art networks, when classifying among the original 10 classes, and was consistently better when transferred to the detection task. One possible explanation for why the preclassifier improves the VGG is that it is able to learn how to best condense the information in the optics-free image.

The base model is trained with a learning rate 0.001 over 45 epochs, while the binary classifier uses a smaller learning rate of $10^{-5}$, and is only trained for 20 epochs. Both models are trained with an Adam optimizer with batch size of 16, and weight decay of $10^{-5}$. The base-class pretraining accuracy, where we train the classifier to predict all 10 classes is shown in Fig. 3a.

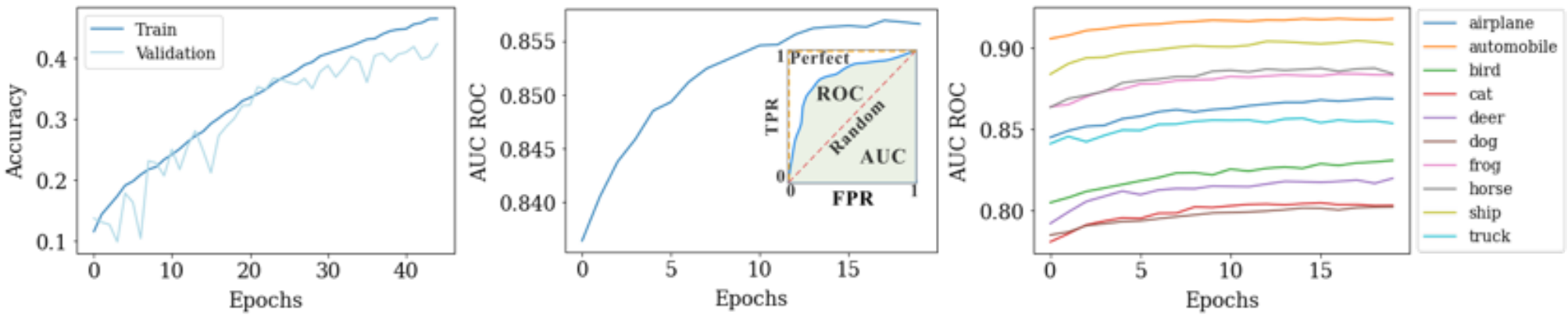

Fig. 3: Results for the detection problem at t=1mm and color Cifar-10 images. (a) Classification accuracy for all 10 classes during pre-training and validation. (b) AUC ROC averaged over all 10 classes. Inset shows an example ROC. (c) AUC ROC for the individual classes.

While accuracy is useful in many classification tasks, it is a poor metric when class sizes are heavily imbalanced (our detection problem) [12]. Therefore, for the detection problem, we use the metric as the area under the curve (AUC) of the receiver operating characteristics curve (ROC). The ROC curve (see inset in Fig. 3b) plots true positive rate (TPR) over false positive rate (FPR) at different classification thresholds, showing the tradeoff between correct positive examples to incorrect negative examples [13], where $\mathrm{TPR} = \frac{\text{true positive}}{\text{true positive} + \text{false negative}}$ and $\mathrm{FPR} = \frac{\text{false positive}}{\text{false positive} + \text{true negative}}$. The AUC metric computes the area under the ROC curve. The area can be interpreted as the network's ability to separate the positive class from the negative classes. An ROC AUC of 0.5 means the classifier cannot separate the classes at all, while a ROC AUC of 1 is perfect separation. We compare our results against a state of the art deep transfer learning based approach, [14] where models trained on ImageNet [15] were transferred to the CIFAR-10 detection problem using the ground-truth CIFAR-10 images. The AUC ROC averaged over all 10 classes for the best model (DenseNet-169) reached 0.997, while ours was 0.85. Considering the loss in the Shannon entropy (ranging from 12% to 19%), this performance in the optics-free images are quite remarkable. To illustrate this point, we showcase exemplary ground truth and sensor images where the detection was successful in Fig. 4.

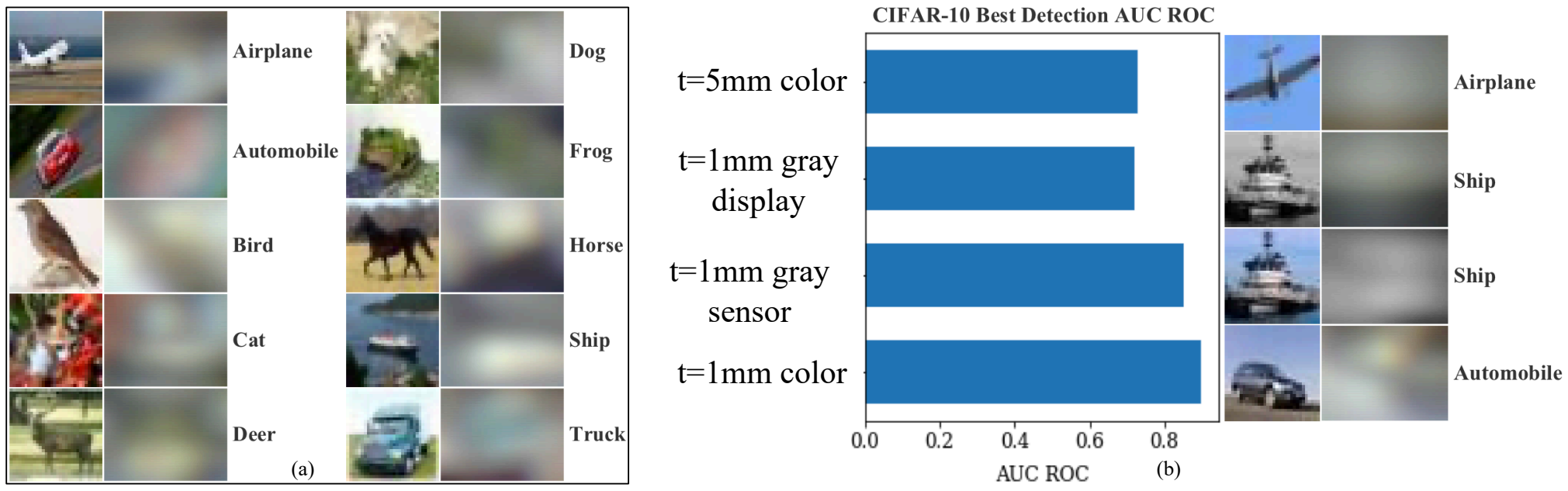

Fig. 4: (a) Exemplary images (left column: ground truth and right column: sensor images) from the Cifar-10 dataset where detection was successfully achieved. (b) Performance summary. Exemplary images from each case are shown to the right.

We further explored the impact of color by using a grayscale version of the Cifar-10 dataset as ground truth. The grayscale conversion was done via the Python PIL library, which uses the ITU-R 601-2 luma transform. All 3 color channels from the image sensor were used as input for the DNN. Training, validation and testing were performed in the same manner as before. The AUC ROC averaged over 10 classes was 0.72. Next, we repeated this experiment by using the original color ground truth images, but by converting the captured sensor images to grayscale using the same PIL library. This approach reached average AUC ROC of 0.85. This difference is interesting and we attribute this to the differences in the loss of information due to the color-to-grayscale conversion process (14% to 19% on average).

Next, we repeated the experiment with t=5mm to elucidate its influence. The only difference was that the base model was only trained for 30 iterations due to overfitting, and no weight decay was used. The color and grayscale images both showed a similar decrease in ROC AUC. The Shannon information loss was estimated as 19%. Previous work on image reconstructions had already confirmed that results were worse at larger t, [6] which is consistent with our results. The average results are summarized in Fig. 4b along with corresponding exemplary images that were correctly detected.

Lastly, we tested our system in a binary setting of fine-grained, balanced classes with the Dogs vs. Cats color dataset with t=1mm. [15] The dataset, which contains 12,500 images of dogs and 12,500 images of cats, was released in a well-known Kaggle competition, and is a subset of images originally used as a CAPTCHA. All images were resized to 20x20 pixels, same as before. The average Shannon information loss for this set was 14%. The 25,000 images were split into 22,000 for training, 2,500 for validation, and 500 for testing. The labels were one-hot-encoded. Here we used the ResNet-50 [16] architecture, which employs identity shortcut connections that allow for the training of deep networks without vanishing gradients. We started with the Keras implementation of the architecture with global max pooling and without the top fully-connected softmax layer. The network ended with the same final 3 layers as in the previous detection (Fig. 2), with an additional neuron in the last layer to match the one-hot encoding of the labels. The model was trained for 70 epochs with an Adam optimizer and learning rate of 0.001. Our approach achieved an average classification accuracy of 92% (Fig. 5). In comparison, Pierre Sermanet, the winner of the Dogs vs. Cats competition attained a classification accuracy of 98.914% on the ground-truth images. This is again remarkable considering (1) the drastic reduction of camera thickness with no optics; and (2) the concomitant loss of Shannon information content (14%).

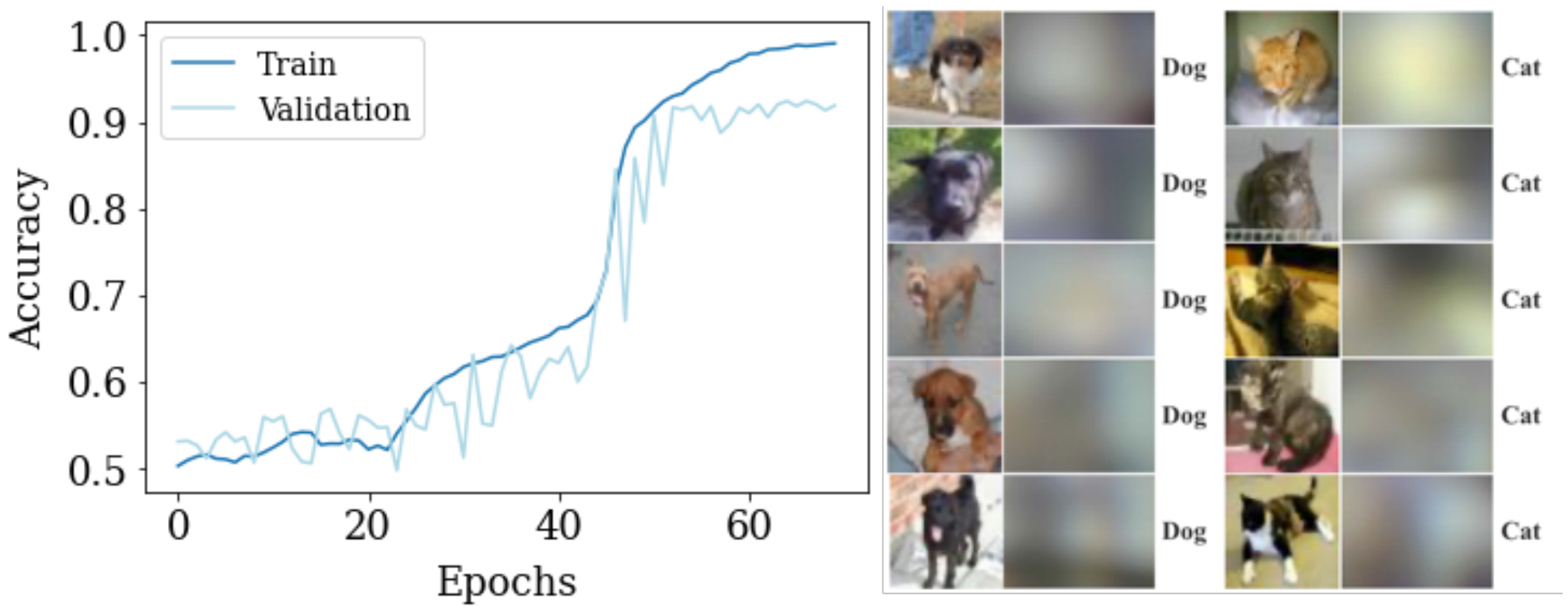

Fig. 5: Left: Classification accuracy for the cats vs. dogs problem. Right: Exemplary images that were successfully classified.

In conclusion, we showed that a deep neural network enabled optics-free camera can classify objects in natural scenes bypassing any human-interpretable image reconstructions. Non-human (machine) imaging is fast becoming the standard use-case for most cameras. In application-specific imaging (our example of cats vs dogs here), the performance of our optics-free cameras can get close to and possibly exceed [7] those of optics-based cameras.

## Funding, acknowledgments, and disclosures

### *Funding*

National Science Foundation (NSF) (1533611). University of Utah Undergraduate Research Opportunities Program (UROP).

### *Acknowledgments*

We would like to thank E. Scullion, Z. Pan and R. Guo for fruitful discussion, and assistance with experiments and software.

### *Disclosures*

RM: University of Utah (P).